\documentclass{ws-rv9x6}
\usepackage{subfigure}     
\usepackage{ws-rv-van}     
\usepackage{graphicx}      
\usepackage{color}         
\usepackage{bm}            
\usepackage{rotating}      
\usepackage{amsmath}       
\makeindex

\begin{document}
\newcommand{\be}{\begin{equation}}
\newcommand{\ee}{\end{equation}}
\newcommand{\bea}{\begin{eqnarray}}
\newcommand{\eea}{\end{eqnarray}}
\newcommand{\eqn}{Eqn.}
\newcommand{\Eqn}{Eqn.}
\newcommand{\tpr}{t^\prime}
\newcommand{\xpr}{x^\prime}
\newcommand{\efpr}{\mathbf{E}^\prime}
\newcommand{\bfpr}{\mathbf{B}^\prime}
\newcommand{\refl}{\mathfrak{r}}
\newcommand{\trans}{\mathfrak{t}}
\newcommand{\tim}[1]{\marginpar{\tiny{\begin{sideways} \sf \textcolor{red}{TF: #1} \end{sideways}}}}

\chapter[Mirror-mediated cooling]{Mirror-mediated cooling: a paradigm for particle cooling via the retarded dipole force}
\label{ra_ch1}              

\author[T. Freegarde and J. Bateman]{Tim Freegarde\footnote{\url{tim.freegarde@soton.ac.uk}} and James Bateman\footnote{Now at the Optoelectronics Research Centre, University of Southampton, Highfield, Southampton SO17 1BJ, U.K.}}
\address{School of Physics and Astronomy, University of Southampton, \\ Highfield, Southampton SO17 1BJ, United Kingdom}

\author[T. Freegarde {\em et al}]{Andr\'{e} Xuereb}
\address{
School of Mathematics \& Physics, The Queen's University of Belfast, \\ Belfast BT7 1NN, United Kingdom}

\author[T. Freegarde {\em et al}]{Peter Horak}
\address{Optoelectronics Research Centre, University of Southampton, \\ Highfield, Southampton SO17 1BJ, United Kingdom}

\begin{abstract}
Cooling forces result from the retarded dipole interaction between an illuminated particle and its reflection. For a one-dimensional example, we find cooling times of milliseconds and limiting temperatures in the millikelvin range. The force, which may be considered the prototype for cavity-mediated cooling, may be enhanced by plasmon and geometric resonances at the mirror.
\end{abstract}

\body

\section{Introduction}\label{ra_sec1}
The atomic physics revolution brought by the Doppler cooling of atoms\cite{haensch75} and ions\cite{wineland75a}, the magneto-optical trap\cite{raab87}, and sub-Doppler cooling\cite{lett 88,dalibard 89,ungar 89}, has prompted the proposal of further schemes\cite{zaugg94,soeding97,horak97,hechenblaikner98,horak98,vuletic00,chan03,kruse03,elsaesser03} to extend the ultracold domain to a wider range of species. Unlike the early reliance upon the scattering force of resonant radiation, these later schemes have largely used the optical dipole force\cite{ashkin70,ashkin78} that results from a spatially-varying interaction with off-resonant illumination. A powerful and flexible means of tweezing atoms and particles\cite{dholakia08}, the dipole force is at heart conservative and incapable of cooling; when coupled to a dissipative component or otherwise invested with a non-Markovian character, however, it too can form the basis of a cooling mechanism\cite{braginskii67,braginskii70}.

\section{Memory in optical cooling}
\label{memory}
Dissipation is intrinsically associated with some form of memory that allows the history of an atom's position --- and thereby its velocity --- to determine its interaction with the optical field. An atom of mass $m$ that moves slowly with velocity $\mathbf{v}$ through a steady optical field with which its interaction is at all times effectively instantaneous will evolve adiabatically and experience a conservative, non-dissipative interaction potential, as is the case for example with the optical dipole force\cite{ashkin70}. If, however, the force $\mathbf{F}$ exerted upon the atom shows a retarded dependence upon the atom's position, then its equation of motion may be written as
\begin{eqnarray}
m \frac{\mathrm{d}^{2}\mathbf{r}}{\mathrm{d}t^{2}} & = & \mathbf{F} \!\! \left( \mathbf{r}(t-\tau)\right) \nonumber \\
& = & \mathbf{F} \!\! \left( \mathbf{r}(t) - \tau \mathbf{v}\right) \nonumber \\
& = & \mathbf{F} \!\! \left( \mathbf{r}(t)\right) - \tau \mathbf{v} \cdot \nabla \mathbf{F} \, ,
\end{eqnarray}
and the force shows a dissipative dependence upon velocity. Provided that the sign is such as to oppose the atom's motion, a cooling mechanism results.

A simple analysis is instructive. In an electric field
\begin{equation}
\mathbf{E}\equiv \mathbf{E}_{0}(\mathbf{r}) \exp i(\omega t - \mathbf{k}\cdot\mathbf{r})
\label{memory1}
\end{equation}
that induces an atomic polarization $\mathbf{P}\equiv \alpha \mathbf{E}$, where $\mathbf{r}$ is the position at time $t$, $\omega$ and $\mathbf{k}$ are the frequency and wavevector of the optical field and $\alpha$ is the atomic polarizability, the instantaneous force $\mathbf{F}$ exerted upon the atom may be written as the gradient of the classical interaction energy $\mathcal{E}$,
\begin{eqnarray}
\mathbf{F} = \nabla \mathcal{E} & = & -\nabla \, \Re (\mathbf{P}\cdot\mathbf{E}^{*})/2 \nonumber \\
& = & -\Re(\alpha \mathbf{E} \cdot \nabla \mathbf{E}^{*})/2 \, .
\end{eqnarray}
Substituting the harmonic optical field of \eqn~(\ref{memory1}) gives
\begin{equation}
\mathbf{F} = \Re(ik\alpha) \mathbf{E}_{0}\cdot\mathbf{E}_{0} + \epsilon_{0} \Re(\alpha) \nabla \mathbf{E}_{0}^{*}\cdot\mathbf{E}_{0} \, .
\label{memory2}
\end{equation}
We shall see that both terms in \eqn~(\ref{memory2}) can result in dissipation, and indeed that both are associated with well-known cooling mechanisms.

\subsection{Doppler cooling}
The first term in \eqn~(\ref{memory2}) corresponds to the scattering force of a near-resonant optical field, which is given a velocity dependence by virtue of the variation of the imaginary part of the atomic polarizability around resonance. Writing the atomic response to a harmonic field at time $t$ in terms of the time-dependent memory of a classical resonator of resonant frequency $\omega_{0}$, decay rate $\Gamma$ and `instantaneous' polarizability $\alpha_{0}$ as
\begin{equation}
\mathbf{P}(t) = \int_{-\infty}^{t} \!\!\!\! \mathbf{E}_{0} \exp \left[ i(\omega t' - \mathbf{k}\cdot\mathbf{r}(t'))\right] \, \alpha_{0} \exp \left[ -i \omega_{0} (t-t')\right] \exp \left[-\Gamma(t-t')\right] \, \mathrm{d}t'
\label{memory3}
\end{equation}
and writing $\mathbf{r} = \mathbf{v} t$ for an atom moving with a velocity $\mathbf{v}$ that has a component $v$ parallel to the wavevector, we obtain, for a uniform field strength,
\begin{equation}
\mathbf{P}(t) = \alpha_{0} \, \mathbf{E}_{0} \exp \left[ i \omega_{0} t\right] \int_{-\infty}^{t} \exp \left[ i\left( \omega - \omega_{0} - \omega \frac{v}{c}\right) t'\right] \exp \left[-\Gamma(t-t')\right] \, \mathrm{d}t'
\end{equation}
and hence, performing the integral and substituting for the electric field at the current atomic position $\mathbf{E}(\mathbf{r}(t))$,
\begin{equation}
\Re(i k \alpha) = \frac{k \alpha_{0} \left[ (1-\frac{v}{c})\omega - \omega_{0}\right] }{\left[ (1-\frac{v}{c})\omega - \omega_{0}\right]^{2} + \Gamma^{2}} \, .
\end{equation}
Within the Lorentzian lineshape of the resonant interaction, the scattering force hence shows a velocity dependence that accounts for the Doppler cooling force\cite{haensch75}.

\subsection{Sisyphus cooling}
The second term in \eqn~(\ref{memory2}) corresponds to the dipole force of an off-resonant optical field,
\begin{eqnarray}
\mathbf{F} & = & \Re(\alpha) \, \nabla \mathbf{E}_{0}^{*} \cdot \mathbf{E}_{0} \nonumber \\
& = & \frac{1}{2} \, \Re(\alpha) \, \nabla |\mathbf{E}_{0}^{2} | \, .
\end{eqnarray}
Although the term $\Re(\chi)$, derived as above for a classical dipole oscillator, shows no first-order dependence upon velocity, such a dependence is indeed introduced when, for example, the polarizability depends upon the relative populations of quantum states between which optical pumping depends upon the instantaneous electric field strength and the accrued population transfer hence depends upon its temporal integral over the atom's recent history, introducing a memory effect as outlined above.

An atom moving within a linearly-polarized, blue-detuned standing wave, for example, undergoes Sisyphus cooling\cite{dalibard85} because the sign of the polarizability differs between the two dressed states whose relative populations are determined by the motion of the atom through the field. For a slowly moving atom, the dressed state population $\Pi(\mathbf{r})$ is given in terms of its steady-state value $\Pi^{\mathrm{st}}(\mathbf{r})$ by\cite{dalibard85}
\begin{equation}
\Pi(\mathbf{r}) = \Pi^{\mathrm{st}} \left( \mathbf{r}-\mathbf{v}/\Gamma_{\mathrm{pop}} \right) \, ,
\end{equation}
where $\Gamma_{\mathrm{pop}}$ is the relaxation rate due to spontaneous emission from the dressed states. A similar mechanism occurs for magnetic atoms within a magnetic field, whereby optical pumping between magnetic sub-levels causes a retardation of the atomic polarization as it moves through an optical polarization gradient\cite{dalibard89,ungar89}, and in three dimensions for atoms that move through the intensity and phase variations of an optical speckle field\cite{horak98}.

\subsection{Cavity-mediated cooling}
For the examples considered above, the retardation upon which cooling depends is due to the memory of the atom as it moves through an unperturbed optical field; the cooling mechanism in each case depends upon spontaneous emission, and its strength and limiting temperature are determined by the spontaneous lifetime. It is also possible, however, for the memory to be held within the optical field when that is modified by its interaction with the atom to provide a retarded back-action upon the atom that perturbed it. A notable example is cavity-mediated cooling\cite{horak97,hechenblaikner98,vuletic00,chan03,kruse03,elsaesser03} whereby an atom or ensemble is not only manipulated by, but also affects, the field within an optical cavity. The retarded dependence of the atom-field interaction upon the atom's position again gives a dissipative, velocity-dependent component to the otherwise conservative trapping force but, rather than depending upon relaxation of atomic state populations, the retardation occurs through decay of the optical field to which the atom is coupled, and it is the cavity decay time that is the significant parameter. The atom and cavity, which separately behave as quantum oscillators with damping rates $\Gamma$ and $\kappa$, are considered to be coupled to form a single quantum system that can be cooled by dissipation in either part.

An extraordinarily wide range of configurations has been considered\cite{domokos03}, with single and multi-mode, linear and ring cavities, longitudinally and transversely pumped to give longitudinal and transverse cooling under strong and weak coupling. Cavity losses can be high or low, and the relative detunings of the atom, cavity and pump laser take most combinations and can resolve or not resolve the motional sidebands of the atomic transition. Cavity-mediated cooling has been considered for individual atoms and ensembles, molecules, quantum-degenerate gases, and macroscopic particles; it is associated with bistable self-organization in one\cite{elsaesser04} and two\cite{baumann10} dimensions; and its optical properties are manifest as a coherent atomic recoil laser (CARL)\cite{bonifacio94a,nagorny03}. As well as cooling the motion of a particle within a rigid optical cavity, it is also possible to cool the motion of the mirrors forming the cavity itself\cite{braginskii70,cohadon99}.

\section{Mirror-mediated cooling}
\label{sect:mmc}
Motivated by a desire to explore the fundamental mechanisms of cavity-mediated cooling without the complexities of its many specific configurations, we have considered a simple geometry whereby the perturbed optical field is returned to the atom by a single mirror. This arrangement differs from conventional cavity-mediated cooling in several respects. Firstly, the nearly-closed system of a resonant cavity, with discrete modes and a characteristic finesse or storage time, is replaced by an open system giving only a single reflection: mirror-mediated cooling is therefore not simply the extreme bad-cavity limit. Secondly, as well as the axial cooling usually considered, we find significant transverse cooling mechanisms, even when there is translational symmetry. Thirdly, while a single mirror cannot offer a cavity's intrinsic amplification of the field response\cite{freegarde02}, we find that cooling may be enhanced by resonances within the mirror, to reach sub-millikelvin temperatures at significant cooling rates. Fourthly, whereas a tightly-focused cavity limits cooling to regions a few wavelengths wide, single mirrors may be replicated as arrays to exploit micron-scale plasmonic resonances and provide cooling over extended distances without cavity alignment or stabilization.

While the absence of the cavity leaves mirror-mediated cooling impractically weak, it is open to a variety of alternative enhancements which may render it both a practical means of cooling cryogenic gas-phase molecules to the millikelvin regime and a significant mechanism by which to influence the dynamics of optically-manipulated macroscopic particles. In contrast to the Casimir-Polder, retarded van der Waals force between a vacuum-polarized particle and its reflection\cite{casimir48a}, it results from the small component of a much larger laser-induced force; unlike quantum friction\cite{pendry97} it needs no resistivity in the mirror; and mirror motion is unimportant\cite{glaetzle09}.


\subsection{Classical description}
\begin{figure}[b]
\centerline{\psfig{file=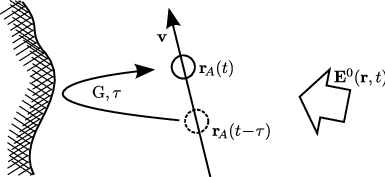}}
\caption{Polarized by incident radiation $\mathbf{E}^{0}$, a particle moving with speed $\mathbf{v}$ is bound to its retarded reflection in a surface, characterized by round-trip time $\tau$ and field propagator $\mathrm{G}$.}
\label{fig:classical}
\end{figure}

The geometry of mirror-mediated cooling is shown schematically in Fig.~\ref{fig:classical}. A particle, moving with velocity $\mathbf{v}$, is illuminated by an incident optical field $\mathbf{E}^{0}(\mathbf{r},t)$ that is scattered and reflected by a nearby surface to encounter the particle once more a time $\tau$ later. The retarded electrostatic interaction between the optically polarized particle and its reflection may, in the Rayleigh-scattering limit, be determined by adapting an analysis of the unretarded binding of point-like particles\cite{depasse94}. We assume an illumination wavelength $\lambda$, particle polarizability $\alpha$ and particle position $\mathbf{r}_{\! A}(t)$, so that the illuminating field experienced by the particle will be $\mathbf{E}^{0}(\mathbf{r}_{\! A}(t))$; the round-trip from the particle to its reflection via the mirror is characterized by a transit time $\tau$ and retarded form of the electric field propagator\cite{haefner09} $\mathrm{G}$ which, like the polarizability and with little loss of generality, we for convenience take here to be scalar. After accounting for multiple interactions, the total electric field $\mathbf{E}(\mathbf{r}_{\! A}(t))$ experienced by the particle at its position $\mathbf{r}_{A}(t)$ at time $t$ proves to be, to lowest order in the velocity $\mathbf{v}$ and $\tau$ and for
$\alpha \mathrm{G} \ll 1$, $\tau |\mathbf{v}| \ll \lambda$,
\be
\mathbf{E}(\mathbf{r}_{\! A}(t)) = \left( 1 - \frac{\alpha \mathrm{G}
\tau}{1-\alpha\mathrm{G}} \mathbf{v}\cdot\nabla \right)
\frac{\mathbf{E}^{0}(\mathbf{r}_{\! A}(t))}{1-\alpha\mathrm{G}} \, .
\label{eq:classicalfield}
\ee
The non-retarded result of Ref.~\refcite{depasse94}, for the particle and its reflection, is thus augmented by a further, velocity-dependent, term. The dipole force upon the particle is obtained as in Ref.~\refcite{depasse94}, taking care over the constraints during differentiation. Expansion in orders of $\alpha\mathrm{G}$ reveals the leading terms for a stationary particle to be the dipole force in the unperturbed field, then the interaction between two dipoles induced by the unperturbed field and, to the same order, the force on the polarized particle due to the field propagated from the induced polarization.

It is convenient to express the field propagator $\mathrm{G}(\mathbf{r}_{A}(t),\mathbf{r}_{A}(t\!-\!\tau))$, which depends upon the current and retarded particle positions, in terms of its value $\mathrm{G}^{0}(\mathbf{r}) \equiv \mathrm{G}(\mathbf{r},\mathbf{r})$ and spatial derivative for a stationary particle,
\bea
\mathrm{G}(\mathbf{r}(t),\mathbf{r}(t\!-\!\tau)) & = & \mathrm{G}^{0}(\mathbf{r}(t)) - \tau \mathbf{v}\cdot \nabla_{1} \mathrm{G}(\mathbf{r}_{1},\mathbf{r}_{2}) \nonumber \\
& = & \mathrm{G}^{0}(\mathbf{r}(t)) - (\tau/2) \mathbf{v}\cdot \nabla \mathrm{G}^{0}(\mathbf{r}(t))
\eea
where $\nabla_{1}$ denotes the derivative with respect to $\mathbf{r}_{1}$, evaluated at $\mathbf{r}_{1} \! = \! \mathbf{r}_{2} \! = \! \mathbf{r}$. For the round-trip time $\tau$, a similar approach requires solution of the circular definition
\bea
\tau(\mathbf{r}(t),\mathbf{r}(t\!-\!\tau)) & = & \tau^{0}(\mathbf{r}(t)) - (\tau/2) \mathbf{v}\cdot \nabla \tau^{0}(\mathbf{r}(t)) \nonumber \\
& = & \frac{\tau^{0}(\mathbf{r}(t))}{1+\mathbf{v}\cdot\nabla \tau^{0}(\mathbf{r}(t))/2} \, .
\eea

For a one-dimensional geometry with a cross-sectional area $\sigma_{l} = \pi w^{2}$, where $w$ is the mode waist, the incident illumination combines with its reflection to give an electric field $\mathbf{E}^{0} = \mathbf{E}_{0} \sin k x$, where $k = 2 \pi/\lambda = \omega/c$; $\mathrm{G}^{0} = i k \exp (2i kx) / 2 \epsilon_{0} \sigma_{l}$ for propagation of the negative frequency complex field component from a dipole a distance $x$ in front of the mirror; and $\tau^{0} = 2 x/c$. The force $\mathbf{F}_{x} = \frac{1}{2}\Re \bigl(\alpha\mathbf{E}\tfrac{\partial}{\partial x}\mathbf{E}^{*}\bigr)$ on the moving particle is hence
\begin{eqnarray}
\mathbf{F}_{x} \! & = & \! \frac{1}{4} \alpha \mathbf{E}_{0}^{2} k \! \left[ \sin 2kx + \frac{\alpha k}{\epsilon_{0} \sigma_{l}} \! \left( 1 \! - \! \frac{v}{c} \right) \sin^{2} \! \! k x (4 \cos^{2} k x \! - \! 1) \right.\nonumber \\
&& \hspace{41mm} - \; \frac{\alpha k^{2}}{\epsilon_{0} \sigma_{l}} \tau v \sin 4 k x \Big]\,.
\label{eq:classicalforce}
\end{eqnarray}

\noindent The force thus comprises three terms. The first two are the dipole force exerted by the unperturbed field, and a Doppler-shifted optical binding force between the particle and its reflection; the Doppler shift here changes the wavelengths of the Fourier field components and hence the gradient of their superposition. The second term is supplemented by a further term in $(v/c)$ due to the missing Lorentz component recently noted in Ref.~\refcite{hinds09}. The third term, which depends upon the particle velocity, the electric field propagator and the round-trip retardation time, is the velocity-dependent force, and dominates the velocity-dependent part of the second term when the distance from the mirror is many wavelengths. Writing the velocity-dependent force as $F_v(x) = -\varrho v$, we thus obtain the friction coefficient
\be
\varrho = \frac{\alpha^{2} \mathbf{E}_{0}^{2} k^{3} \tau}{4 \epsilon_{0} \sigma_{l}} \sin 4 k x \, .
\label{eq:classicalfriction}
\ee
When the sign of this component is such as to oppose the particle velocity, cooling ensues.

\subsection{Semi-classical description}
\Eqn\ (\ref{eq:classicalfriction}) may be reproduced by a semi-classical analysis\cite{xuereb09b} of the cooling of a two-level atom, of linewidth $\Gamma$, that is coupled to the mirror by a single transverse mode of the electromagnetic field\cite{xuereb09b}. This one-dimensional geometry resembles an experimental arrangement that has been considered in detail\cite{bushev04,rist08,hetet10}, but we consider only a single transverse electromagnetic mode corresponding, for example, to the inclusion of a spatial filter or the use of an optical fibre instead of the free-space delay line. The model thus consists of a single two-level atom coupled to the modes of the electromagnetic field in the right half space, according to the Hamiltonian
\begin{eqnarray}
H & = & \hbar \omega_{a} \sigma^{+} \sigma^{-} + \frac{p^{2}}{2m} + \int \hbar \omega a^{\dagger}(\omega) a(\omega) \, \mathrm{d}\omega \nonumber \\
&& - i \hbar g \int \sin \left( x\frac{\omega}{c} \right) \left[ \sigma^{+} a(\omega) - a^{\dagger}(\omega) \sigma^{+} \right] \, \mathrm{d}\omega \, .
\label{hamiltonian}
\end{eqnarray}
The first line represents the internal atomic energy given by the transition frequency $\omega_{a}$, the kinetic energy, and the mode energies, respectively, and the second line is the interaction energy between the atomic dipole and the electromagnetic modes with a coupling coefficient $g$ which is assumed constant over the narrow range of relevant frequencies. The density operator $\rho$ of the atom-field system follows the master equation
\begin{equation}
\frac{\mathrm{d}\rho}{\mathrm{d}t} = - \frac{i}{\hbar} \left[ H, \rho \right] + \mathcal{L} \rho
\label{master}
\end{equation}
where $\mathcal{L}\rho$ is a standard Liouville term corresponding to spontaneous atomic decay into free space modes with rate $\Gamma$. Eqns.~\ref{hamiltonian} and \ref{master} are simplified by adiabatic elimination of the atomic excited state. Treating the atomic momentum $p$ and position $x$ as classical variables and deriving the Heisenberg equations of motion for the mode operators $a(\omega)$, which are then approximated by complex numbers with initial distribution $a(\omega) = A \delta(\omega-\omega_{0})$, the stationary solution is then found to lowest order in the atom-field coupling $g$, yielding the friction coefficient
\be
 \varrho = \hbar k^2 \Gamma\tau s \frac{\sigma_a}{2\sigma_{l}}\sin(4kx) \, ,
 \label{eq:semiclassicalfriction}
 \ee
where the atomic scattering cross-section $\sigma_a = 3\lambda^2/(2\pi)$, we use $2\pi g^2 = \Gamma\sigma_a/\sigma_{l}$ \cite{domokos02}, and $s = g^2|A|^2/\Delta^2$ is the atomic saturation assuming the pump detuning $|\Delta| \equiv |\omega_0-\omega_a|\gg\Gamma$. The substitutions $\alpha \equiv \pi \epsilon_{0} \sigma_{l} g^{2} / k\Delta$ and $\pi \sigma_{l} c \epsilon_0 \mathbf{E}_{0}^{2} \equiv 4 \hbar \omega A^{2}$ return the classical result of Eq.\ (\ref{eq:classicalfriction}).

\begin{figure}
\centerline{\psfig{file=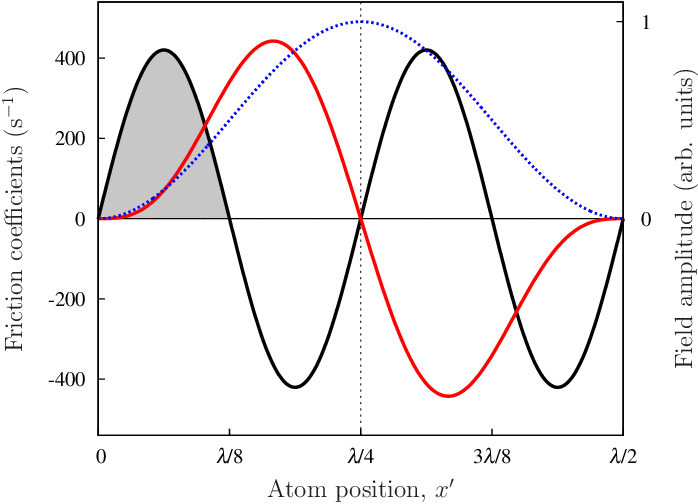,width=7cm}}
\caption{Spatial dependence of the lowest order longitudinal (black) and transverse (red) friction coefficients $\varrho/m$ for $s=0.1$, $\sigma_a/(\pi w^2)=0.1$, $x=4.0$ m. Shading indicates the region in which cooling occurs in both dimensions. The dotted blue line shows the intensity of the pump wave (arbitrary units). The atom position $x^\prime$ is measured from the nearest field node.}
\label{fig:friction}
\end{figure}

Fig.\ \ref{fig:friction} shows how the friction varies with atomic position in the standing pump wave for the example of $^{85}\mathrm{Rb}$. Regions of cooling ($\varrho>0$) alternate with regions of heating ($\varrho<0$) and, for the chosen parameters, predicted cooling times $(m/\varrho)$ are a few ms at points of greatest friction. The same geometry is also predicted to show friction in the transverse direction: the coupling constant $g$ is a function of the particle's radial position, and cooling results when the temporal intensity variation causes the deceleration as the particle leaves the beam to exceed the acceleration as it enters. When the mode waist is comparable with the optical wavelength we find the transverse and longitudinal cooling forces to be similar in magnitude\cite{xuereb09b}. Although the two components differ in their spatial dependence, regions exist in which the particle is cooled simultaneously in both directions.

We estimate the steady-state temperature achievable with this cooling scheme by applying the fluctuation-dissipation theorem\cite{xuereb12} and using the well-known result for diffusion in the standing wave of the pump field\cite{gordon80}. To leading order and in the weak coupling limit assumed above, this leads to a stationary temperature at the position of maximum friction given by
\be
k_{B} T = \frac{D}{\varrho} \approx \frac{\hbar}{\tau} \frac{\sigma_{l}}{\sigma_{a}} \, ,
\label{eq:temp}
\ee
where the diffusion coefficient $D$ is the same as for Doppler cooling. The stationary temperature therefore resembles the Doppler limit, with the atom-mirror round-trip time $\tau$ replacing the atomic coherence time $1/\Gamma$, scaled by the ratio of the cross-section $\sigma_{l}$ of the pump beam to that of the atom, and is independent of detuning, pump intensity and atomic saturation. For atomic rubidium with $x=4.0$~m and the same parameters as in Fig.\ \ref{fig:friction}, we obtain a temperature of 380~$\mu$K.

The alternation with position between cooling and heating means that, for a net cooling effect, particles must be confined to within $\lambda/8$, e.g., by a far-detuned dipole trap\cite{ritt06}. Such confinement couples the mean kinetic energy of the particle to an equal but uncooled mean potential energy, halving the cooling rate. Our analytical expression Eq.\  (\ref{eq:temp}) must be modified to account for the fraction of the particle's trajectory, lasting of order $\tau$ after each apogee of the trapped motion, during which the (newly-reversed) motion is accelerated rather than diminished by the retarded force. There is a further small correction because the trapped particles' trajectories extend beyond the region of maximal cooling. For the parameters of Fig.\ \ref{fig:friction} with a trap frequency $\nu_\textrm{\footnotesize trap} = 1.5$~MHz, these corrections increase the temperature to 580~$\mu$K. While this is greater than the Doppler temperature of 140~$\mu$K, the independence from detuning means that, for sufficiently off-resonant operation ($|\Delta|\gtrsim 10\Gamma$), mirror-mediated cooling will be the dominant cooling effect\cite{xuereb09b}.

A full three-dimensional treatment of the situation is also possible using the methods outlined in Sect.~\ref{memory} above. Under circularly-polarized illumination and taking into account the `near-field' terms in the full electric dipole tensor propagator $\mathbf{G}$, which depend upon $1/R^{2}$ and $1/R^{3}$ where $R$ is the dipole-dipole separation, we find that a non-zero friction force
\begin{equation}
\left<\left(\begin{array}{c}
  F_x\\
  F_y\\
  F_z
\end{array}\right)\right>=(-v)\frac{\alpha^2 \mathbf{E}_0^2 k^2}{32\pi \epsilon_0 c^{3} \tau^2} \left(\begin{array}{c}
  1\\
  3\\
  3
\end{array}\right)
\end{equation}
remains after oscillating terms have been eliminated by averaging over the $x$ coordinate. This force becomes comparable with the amplitude of the position-dependent force when $x\lesssim\lambda$, and could therefore be particularly significant for refractive macroscopic particles in suspension.

We envisage that initial experimental demonstration at the atomic level might best be carried out with trapped ions, which may be readily localized to of order 10~nm\cite{bushev04}. Full confinement to the cooling regions may be unnecessary provided that particles spend more time being cooled than heated. For transverse cooling of a molecular beam, for example, it may suffice to channel the molecules using the periodic potential of an additional standing wave or bichromatic evanescent field.

The semi-classical approach allows an accurate description of the quantum fluctuations of both photon number and atomic momentum, and is valid for arbitrary coupling strength and atomic velocity. It readily yields the limiting temperature at which cooling is balanced by heating from off-resonant scattering, and we have confirmed our analytical results with Monte--Carlo simulations of the dynamics of a rubidium atom coupled to a discretized set of modes \cite{horak01b} in the presence of an external harmonic trap. It is not however easily extended to more complex geometries in three dimensions or to the damping of extended\cite{chaumet01} or highly polarizable particles; for these we find our original classical approach, with its arbitrary forms of $\tau$ and $\mathrm{G}$, to be more instructive.

\section{Binding and higher order cooling terms}
The friction term of Eqns.~(\ref{eq:classicalfriction},\ref{eq:semiclassicalfriction}) is the leading term in a series expansion of the interaction between the illuminated particle and its retarded reflection, and corresponds to the sum of two situations: that of a particle, polarized by the unperturbed illumination, in the once-scattered field; and that of a particle, polarized by the once-scattered field, in the unperturbed illumination. In the one-dimensional case considered above, the friction force derives from the velocity-dependence of the relative phase between these two fields (where the unperturbed illumination field has two wavevector components because it is itself reflected by the mirror); in three dimensions, there is also a contribution from the velocity-dependence of the field magnitudes.

Subsequent terms in the series expansion can also be significant, however, and it is instructive to label contributions to the field according to the number of times it has been scattered by the particle. We first split the standing-wave field $\mathbf{E}^{0}$ into travelling components $\mathbf{E}_{+}^{0}$ and $\mathbf{E}_{-}^{0}$, and now let $\mathbf{E}_{\pm}^{1}$ be the field, derived from $\mathbf{E}_{\pm}^{0}$, which arrives back at the particle following a single scattering process and round-trip to the mirror and back. In other words,
\be
\mathbf{E}_{\pm}^{1} = G \alpha \mathbf{E}_{\pm}^{0} \, .
\ee
This field will in turn polarize the particle, leading, after another round-trip, to a further field $\mathbf{E}_{\pm}^{2}$, and so on. The total energy of the polarized particle in the overall field will therefore be
\be
\mathcal{E} = \frac{1}{2} \sum_{i,j,m,n} \mathbf{P}_{i}^{m} \cdot \mathbf{E}_{j}^{n}
\ee
where the polarization component $\mathbf{P}_{i}^{m} = \alpha \mathbf{E}_{i}^{m}$ is induced by the $i$th component of the unperturbed illumination after it has been scattered and reflected $m$ times. It will be apparent that terms with the same $m+n$ have the same dependence upon $G$ and $\alpha$. When $m=n=0$, we have the potential for the particle due to the usual dipole tweezing force upon a polarizable particle in the unperturbed laser beam, while the friction terms of Eqns.~(\ref{eq:classicalfriction},\ref{eq:semiclassicalfriction}) are given by combining the cases $\mathbf{P}_{\pm}^{0} \cdot \mathbf{E}_{\mp}^{1}$ and $\mathbf{P}_{\pm}^{1} \cdot \mathbf{E}_{\mp}^{0}$.

\begin{figure}[b]
\centerline{\psfig{file=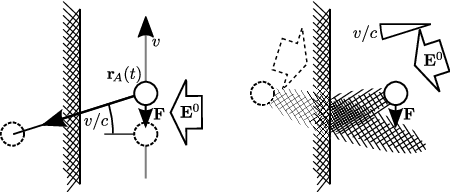,width=0.6\textwidth}}
\caption{Retarded binding of a normally-illuminated particle, moving with speed $v$, to its own reflection, depicted (a) in the laboratory frame, in which the image lags behind; (b) in the rest frame of particle, whereby the `wake' trails behind.}
\label{fig:binding}
\end{figure}

Terms for which $m=n$, whereby the paths taken by the two field components are identical, avoid the acute spatial dependence that stems from interference between different routes from the illumination source to the particle. The term $\mathbf{P}_{\pm}^{1} \cdot \mathbf{E}_{\pm}^{1}$ is of particular significance, for it corresponds to the tweezing force upon the particle due to its reflected scattered field. Within the `shadow' of a refractive particle, the field is dominated by the scattered field, just as it is for a glass lens, and the friction term may be considered to be a component of the optical binding force\cite{burns89,metzger06a} between the particle and its delayed reflection. An example is the situation shown in Fig.\ \ref{fig:binding}(a), in which a highly refractive particle moves parallel to a plane mirror under perpendicular illumination. The finite time taken for light from the particle to return via the mirror causes the reflected image to trail behind the moving particle, providing a component of the binding force in the direction of the particle velocity and therefore a transverse force even when the geometry shows translational symmetry. This interaction, between the propagated field and the further polarization that it induces, requires the consideration of the vector nature of the electromagnetic field and higher order terms than were retained in \Eqn\ (\ref{eq:classicalforce}), without which the translational symmetry would render it identically zero. Fig.\ \ref{fig:binding}(b) shows the same geometry in the rest frame of the particle, in which it is the inclination of the transformed illumination that causes the focused `wake' again to lie behind the particle.

The friction force of Fig.\ \ref{fig:binding} is significant only when the particle is strongly coupled to its reflection and, for the planar geometry illustrated, requires highly polarizable particles. It again alternates with distance from the mirror, but it does so asymmetrically because the apparent field strength described by Eq.\ (\ref{eq:classicalfield}) is also modulated. The result, assuming Rayleigh scattering (no multiple scattering within the particle) and for an illuminating electric field polarized parallel to the atomic velocity in the $y$-direction, is a non-zero spatial average force, which can therefore apply to whole atom clouds or extended particles,
\be
F_{y} = (-v_{y}) \frac{2 \alpha^{3} k^{4} \mathbf{E}_{0}^{2}}{(4\pi\epsilon_{0})^{2} c R^{3}} \, ,
\ee
where $R = c \tau$; if the electric field is in the $z$-direction, the frictional force is a factor of 2 weaker. This force becomes comparable with the amplitude of the position-dependent force when $x \lesssim \sqrt{k \alpha/\epsilon_{0}}$, and could therefore be particularly significant for refractive mesoscopic particles in, for example, colloidal photonic crystals, whereby the reflection is generated by Bragg scattering within the crystal itself; in such cases, however, a Mie scattering calculation is likely to be necessary\cite{Metzger06b}.

\section{Transfer matrices}
The one-dimensional geometry considered in Sect.~\ref{sect:mmc} may be succinctly described through a transfer matrix approach\cite{deutsch95}, which allows the fields interacting with a linear scatterer to be determined in the presence of a series of optical elements; while restricted to a single spatial dimension, this formalism nonetheless allows the analysis of a wealth of interesting physical situations. For the analysis of optical cooling forces, we have extended the transfer matrix model to allow it to account for moving as well as static scatterers\cite{xuereb09a}; the method is more general than an analysis based on modal decomposition, and may be applied to a range of mechanisms including Doppler, polarization gradient, and cavity-mediated cooling, for it is applicable to any scattering object, be it an atom, particle, dielectric slab, or mirror.

\begin{figure}
\centerline{\psfig{file=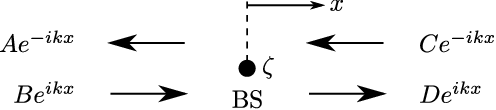,width=0.6\textwidth}}
\caption{The four different modes that interact through a point-like beamsplitter in one dimension.}
\label{fig:transfermatrix}
\end{figure}

The transfer matrix approach begins by expressing the fields either side of each scatterer, i.e. a particle or optical element modelled as a beamsplitter, in terms of a discrete sum of left- and right-propagating plane wave modes with different wavenumbers $k$ and amplitudes $A(k)$, $B(k)$, $C(k)$ and $D(k)$, as shown in Fig.~\ref{fig:transfermatrix}, so that the electric fields are
\begin{equation}
 \label{eq:Efield}
 \mathbf{E}=\begin{cases}
    \sum_k\big[A(k)e^{-ikx-i\omega t}+B(k)e^{ikx-i\omega t}\big]+\rm{c.c.}\\
    \sum_k\big[C(k)e^{ikx-i\omega t}+D(k)e^{-ikx-i\omega t}\big]+\rm{c.c.} \, ,
   \end{cases}
\end{equation}
with $A(k)$ and $B(k)$ being the mode amplitudes on the left side,  $x<x_{\text{BS}}(t)$, while $C(k)$ and $D(k)$ being the amplitudes on the right side, $x>x_{\text{BS}}(t)$, of the beamsplitter $\text{BS}$. In accordance, the magnetic field is~\cite{jackson75}
\begin{equation}
 \label{eq:Bfield}
 c \mathbf{B}=\begin{cases}
    \sum_k\big[-A(k)e^{-ikx-i\omega t}+B(k)e^{ikx-i\omega t}\big]+\rm{c.c.}\\
    \sum_k\big[-C(k)e^{-ikx-i\omega t}+D(k)e^{ikx-i\omega t}\big]+\rm{c.c.} \, .
   \end{cases}
\end{equation}

The role of the transfer matrix $M$ is then to connect the field amplitudes on the right of the scatterer to those on the left. This relation is well-known\cite{deutsch95} for the stationary case; for a moving scatterer, we must first transform the electromagnetic field into a frame moving with the instantaneous velocity $v$ of the scatterer. In this frame, the interaction of the field with the scatterer at $x'=0$ may be characterized by the single scattering strength parameter $\zeta$ by means of the one-dimensional wave equation\cite{jackson75,deutsch95},
\begin{equation}
\label{eq:WaveEqn}
\left(\partial_{\xpr}^2-\frac{1}{c^2}\partial_{\tpr}^2\right) \efpr(\xpr,\tpr) = \frac{2}{kc^2}\zeta\,\delta(\xpr)\,\partial^2_{\tpr} \efpr(\xpr,\tpr)\,.
\end{equation}
The electric field is again addressed by a modal decomposition similar to \Eqn~(\ref{eq:Efield}),
so that the scattering in this frame can be fully described within a closed set of modes,
\begin{equation}
\efpr(x,t) =\begin{cases} A' e^{-ik\xpr -i\omega \tpr} + B' e^{ik\xpr -i \omega \tpr} + c.c. &\xpr < 0\\
C' e^{-ik\xpr -i\omega \tpr} + D' e^{ik\xpr -i \omega \tpr} + c.c. &\xpr > 0 \, ,
\end{cases}
\end{equation}
where the index $k$ has been dropped. The linear relation between the field amplitudes on the right of the scatterer and those on the left can then be derived from the wave \Eqn~(~\ref{eq:WaveEqn}),
\begin{equation}
\label{eq:TM_fix}
\begin{pmatrix}
 C^\prime\\
 D^\prime
\end{pmatrix}=M_0\begin{pmatrix}
 A^\prime\\
 B^\prime
\end{pmatrix}
\end{equation}
with
\begin{equation}
\label{eq:M0}
M_0 = \begin{bmatrix}
 1-i\zeta & -i\zeta\\
 i\zeta & 1+i\zeta
\end{bmatrix}
= \frac{1}{\trans}
\begin{bmatrix}
 1 & -\refl   \\
\refl   & \trans^2 - \refl^2
\end{bmatrix} \, ,
\end{equation}
where the second form of the transfer matrix is expressed in terms of the reflectivity $\refl$ and transmissivity $\trans$ of the beamsplitter. This latter form is more convenient for the description of moving mirrors, whereas for atoms the scattering strength parameter $\zeta$ is readily expressed in terms of its polarizability $\alpha$ by
\begin{equation}
 \zeta = \frac{\pi \alpha}{\epsilon_0 \lambda S}\,,
\end{equation}
where $\alpha$ is the linear polarizability and $S$ is the effective beam cross section. For a two-level, unsaturated atom with transition frequency $\omega_{\text{A}}$ and linewidth $\Gamma$ (FWHM), for example,
\begin{equation}
 \label{eq:ZetaTLA}
 \zeta = \frac{\sigma_{\text{A}}}{2 S} \frac{\Gamma/2}{\omega_{\text{A}}-\omega - i\Gamma/2}\,,
\end{equation}
where $\sigma_{\text{A}}= \tfrac{3\lambda^2}{2\pi}$ is the resonant radiative cross section of an atom. In this case the transfer matrix depends on the wave number $k$, which might lead to significant effects, e.g., Doppler cooling, close to resonance with the atom.

The transformation back into the laboratory frame involves the change of coordinates $x' = x-vt$ and $t'=t$ and the Lorentz-boost of the electric field up to linear order in $v/c$,
\be
\mathbf{E} = \mathbf{E}' + v \mathbf{B} \, .
\ee
The transformation may be expressed in terms of the linear matrix $\hat{L}(v)$ as
\begin{equation}
\begin{pmatrix}
 A(k)\\
 B(k)
\end{pmatrix}=\hat L(-v)\begin{pmatrix}
 A^\prime(k)\\
 B^\prime(k)
\end{pmatrix}
\end{equation}
with
\begin{equation}
\label{eq:Lmatrix}
\hat L(v) = \begin{bmatrix}
 \left(1+ \frac{v}{c}\right) \hat{P}_{-v} & 0\\
0 & \left(1- \frac{v}{c}\right) \hat{P}_v
\end{bmatrix} \, ,
\end{equation}
where the operator $\hat{P}_v:f(k)\mapsto f\left(k+k\tfrac{v}{c}\right)$ represents the Doppler-shift of the plane waves in a moving frame and $\hat L^{-1}(v) = \hat L(-v)$. The total action of the moving beamsplitter
\begin{equation}
\label{eq:ABCD}
\begin{pmatrix}
 C(k)\\
 D(k)
\end{pmatrix}= \hat{M}
\begin{pmatrix}
 A(k)\\
 B(k)
\end{pmatrix}\text{,}
\end{equation}
may then be obtained from
\begin{align}
\label{eq:LML}
 \hat{M} &= \hat L(-v) M_0 \hat L(v) \nonumber \\
 &=\frac{1}{\trans}\begin{bmatrix}
 1 & -(1-2\tfrac{v}{c})\refl\hat{P}_{2v} \\
 (1+2\tfrac{v}{c})\refl\hat{P}_{-2v}& \trans^2-\refl^2
\end{bmatrix} \, ,
\end{align}
where the second line follows if $\refl$ and $\trans$ are taken to be independent of wavenumber. This result differs from the case of the stationary scatterer in \Eqn~(\ref{eq:M0}) in that the off-diagonal terms include the Doppler shift imposed by the reflection from a moving mirror, so that the coupled counter-propagating plane wave modes differ in wave number. If the polarizability itself depends upon the wavenumber $k$, as in the case of Doppler cooling of a moving atom, the Doppler shift operator is taken to act also upon it. The operator $\hat{P}$ can be Taylor-expanded to first order in the parameter $v/c$.

\begin{figure}[t]
\centerline{\psfig{file=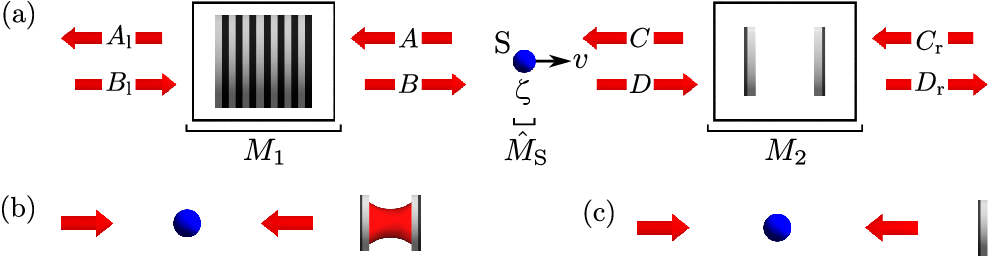,width=0.95\textwidth}}
\caption{(a) An example of an optomechanical system that may be analysed using transfer matrices. Here, a moving atom $S$ lies between to immobile optical elements $M_{1}$ (a Bragg reflector) and $M_{2}$ (a Fabry-P\'{e}rot cavity). Two specific configurations are (b) an atom in front of a two-mirror cavity and (c) the 1-D prototype for mirror-mediated cooling.}
\label{fig:generic}
\end{figure}

A given geometry of fixed and mobile scatterers is then built up by combining transfer matrices for the individual components, as shown schematically in Fig.~\ref{fig:generic} and, in the simple but common case of pumping with monochromatic light, the equations relating the field amplitudes can be solved analytically. This yields closed-form expressions for the static and friction force acting on the mobile scatterer. We emphasize once more that the above expressions hold generically, whatever the magnitude of the polarizability of the scatterer. Further details of the transfer matrix approach to moving scatterers may be found in Refs.~[\refcite{xuereb09a,xuereb10a,xuereb12}].

\section{Enhancements and optomechanics}
The prototype geometries considered so far suffer two weaknesses. Firstly, they share with cavity-mediated cooling schemes the requirement for a tightly focused image to provide the strong coupling of the particle to the field that affects and is affected by it. Cooling therefore occurs only within a small volume. Secondly, they lack the enhancement offered, in cavity-mediated cooling, by the optical resonator. That the atoms no longer need to be enclosed within a resonant cavity, however, avoids a number of experimental difficulties (e.g. if viewports must be enclosed), and the simple geometry of mirror-mediated cooling permits a number of alternative enhancements, which may be analysed using the transfer matrix approach.

\subsection{External cavity cooling}
\begin{figure}
\centerline{\psfig{file=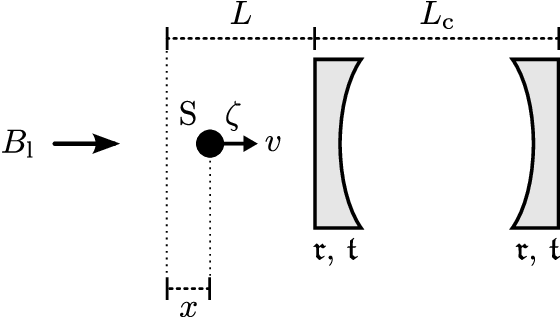,width=0.5\textwidth}}
\caption{External cavity cooling: the use of a resonant cavity in place of a simple mirror may greatly enhance the retardation without the complications associated with enclosing the particle inside the cavity.}
\label{fig:external}
\end{figure}

\begin{figure}
\centerline{\psfig{file=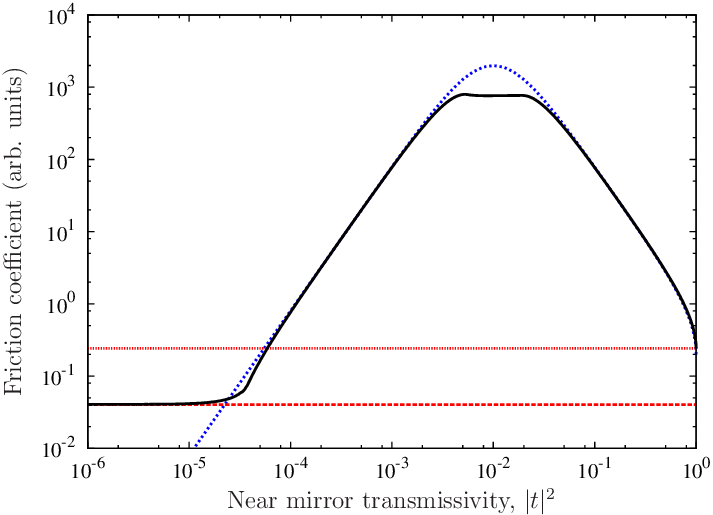,width=0.6\textwidth}}
\caption{The amplitude of the friction acting on a scatterer (black line), with polarizability $\alpha=0.001$, interacting with a cavity tuned to achieve maximum friction, as we vary the transmissivity, $|t|$, of the near mirror. Also shown is the friction force for the scatterer interacting with the near (dashed red line) or the far (dashed--dotted red line) mirror only. The intracavity field is drawn (dashed blue line) as a guide to the eye.}
\label{fig:enhancement}
\end{figure}

Firstly, as shown in Fig.~\ref{fig:external}, the mirror may be replaced by an external resonator\cite{xuereb10b}, which allows the same effective retardation $\tau$ to be achieved with a greatly reduced mirror distance and therefore provides the memory of cavity-mediated cooling but not the enhancement in intensity. Fig.\ \ref{fig:enhancement} shows the calculated friction coefficient when the mirror is combined with a second reflecting surface of transmissivity $t$ to form a Fabry-P\'{e}rot etalon, and tuned to provide maximum friction. We find that the friction force can be enhanced by a factor approaching the cavity finesse until the losses of the combined atom--cavity system are dominated by scattering at the atom. For highly scattering particles such as levitated nanoparticles\cite{barker10,chang10}, external cavity cooling may allow higher enhancement factors than when the particle is placed within the resonator.

While such enhancement could be applied to a planar geometry such as Fig.\ \ref{fig:binding}, we also envisage its use with microfabricated reflectors, in which material or spatial resonances such as the plasmon modes of a micro-structured mirror or antenna might also enhance the strength of the scattered field. As possible examples, we suggest microscopic antennae\cite{smythe07} or concave, plasmon-resonant cavities\cite{sugawara06,kelf06}. Such tiny individual reflectors could then be replicated as arrays, which are practically impossible with cavity-mediated cooling which would require each reflector to be precisely aligned. Like a cobbled street for a bicycle, the dissipative effect would then be reproduced over an extended area. Combined with a planar trap, such as a bichromatic evanescent field\cite{labeyrie96}, to confine the particles above the surface, such geometries could provide a greater integrated cooling effect than the swift passage through a single cooling region.

\subsection{Amplified feedback cooling}
\begin{figure}
\centerline{\psfig{file=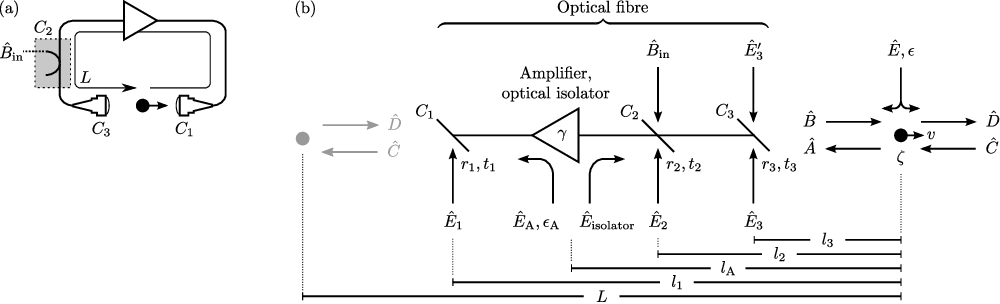,width=\textwidth}}
\caption{Amplified feedback cooling in a unidirectional ring cavity. (a) Schematic arrangement; (b) equivalent, unfolded, transfer matrix model, showing the particle on both sides of the feedback optics to illustrate the recursive nature of the cavity.}
\label{fig:amplified}
\end{figure}

For basic mirror-mediated cooling in three dimensions, there is a trade-off between the strong coupling $G$ between the particle and the optical field for small mirror distances and the increased retardation $\tau$ with a distant mirror. One possibility, illustrated in Fig.~\ref{fig:amplified}(a), is to introduce an optical gain element into the retarded feedback\cite{xuereb11b} to provide a passive version of parametric feedback cooling\cite{gieseler12}. An additional advantage of the ring geometry shown here is that a far-field spatially-averaged cooling force, absent in the prototype linear geometry, can occur for the $\mathbf{P}_{+}^{m} \cdot \mathbf{E}_{+}^{n}$ term with $m+n=1$ because the illumination field co-propagates with the returning scattered field.

After interacting with a scatterer moving in free space, light can be divided into two branches, depicted in Fig.~\ref{fig:amplified}(b):\ the transmitted part, which carries no first-order information about the motion, and the reflected part, which does. Whereas in a Fabry--P\'erot cavity these two branches are invariably mixed, by placing the scatterer inside a ring cavity one can use an optical amplifier and an isolator to amplify the reflected light whilst removing the transmitted light. This amplifies the interaction between light and the motion of the scatterer, and at the same time gets rid of excess noise that does not contribute to cooling but acts to increase momentum diffusion. Cavity cooling of atoms inside a Fabry--P\'erot cavity with a gain medium was considered in 2001 by Vuleti\'c~\cite{vuletic01a}, who noted that the amplification of spontaneous emission represents one limitation of the mechanism. Indeed, in the system discussed by Vuleti\'c, the noise resulting from spontaneous emission in the `forward' direction is also amplified and sent back to the atom.

The asymmetric ring cavity system~\cite{xuereb11b}, schematically shown in Fig.~\ref{fig:amplified}, can be modelled using the transfer matrix method presented above. Increasing the gain of the amplifier narrows the parameter range in which cooling is seen, but acts to significantly improve the cooling rate without a similar increase in the momentum diffusion experienced by the scatterer. Even for a modest gain factor, less than $2$, this leads to minimum temperatures that are significantly lower than would be the case without an amplifier.

\section{Conclusion}

Mirror-mediated cooling is firstly a useful prototype to elucidate the mechanisms of cavity-mediated cooling, which proves in each of its various geometries to provide a delayed back-action, mediated by the dipole force, upon the cooled particle. The introduction of a time delay into an optical system allows the dipole interaction to be endowed with a non-conservative nature. Each geometry may be equivalently regarded as providing dipole-dipole interactions between the illuminated particle and its retarded reflection, and a series expansion of the interaction proves to offer further insight into the cooling mechanisms and their characteristics. For cavity-mediated cooling, the cavity then enhances the fundamental mechanisms by amplifying the field intensity and retardation and by converting variations in phase to variations in intensity.

The cavities of cavity-mediated cooling mechanisms --- and cavity-atom systems in general --- are commonly regarded as resonant quantum oscillators that are conceptually equivalent to the atoms to which they are coupled, and rendered non-ideal through a perturbing loss that corresponds to the spontaneous decay of the atom. Mirror-mediated cooling proves to be more than the `very bad cavity' limit, in that it is characterized by a single recurrence rather than a steady, exponentially-decaying memory, and therefore offers a somewhat different paradigm for the analysis of atom-field interactions.

Although the fundamental process of mirror-mediated cooling tends to be extremely weak, it is open to a variety of enhancements that, in contrast to cavity-mediated cooling, do not require the cooled particle to be wrapped within a high finesse resonator. The cavity, for example, may lie to one side of the particle, allowing higher $Q$-values and more straightforward alignment. The introduction of active gain may allow considerable enhancements in both cooling rate and active volume (and hence longer interaction times with moving particles), as well as the compensation of passive losses.

Because it is based upon the off-resonant dipole force and an open geometry compatible with microfabricated elements\cite{Gigan2006,Groblacher2009a,Eichenfield2009} or dielectric microspheres\cite{Li2010c,RomeroIsart2011}, mirror-mediated cooling and its enhancements may be useful for extending atom cooling techniques to a wider range of atomic and molecular species, including Rydberg atoms and heavy molecules, and mesoscopic particles. Like the dipole force in general, it is likely to be most significant for wavelength-scale objects: the cooling of a nanomechanical oscillator to its ground vibrational state\cite{OConnell2010,Teufel2011,Chan2011}, or of a particle levitated above a photonic crystal\cite{rahmani06} whose distributed resonance provides a long retardation time and strong particle-field coupling.

To explore these interactions, we have outlined an extended formalism based on the transfer matrix method that may be applied to a broad variety of geometries. This common formalism leads to a natural unification of the cooling mechanisms for atomic motion\cite{Kruse2003,Murch2008,Brennecke2008,Purdy2010,Schleier-Smith2011} and optomechanical systems.

\section*{Acknowledgements}
We gratefully acknowledge helpful discussions with P. Domokos, H. Ritsch and C. Zimmermann, and support from the U.K. Engineering \& Physical Sciences Research Council grants EP/E058949/1 and EP/E039839/1 and the European Science Foundation's EuroQUAM project {\em Cavity-Mediated Molecular Cooling}. AX acknowledges funding from the Royal Commission for the Exhibition of 1851.

\bibliographystyle{ws-rv-van}
\bibliography{mirrorcooling}

\printindex                         
\end{document}